\documentstyle[12pt]{article}

\begin{document}
\begin{titlepage}
\setcounter{page}{1}
\title{\bf Cosmological perturbations and classical change of signature}
\author{J\'er\^ome Martin$^*$ \\
\mbox{\small \em Department of Applied Mathematics and Theoretical Physics} \\
\mbox{\small \em University of Cambridge, Silver street} \\
\mbox{\small \em Cambridge CB3 9EW, United Kingdom}}
\maketitle
\begin{abstract}
Cosmological perturbations on a manifold admitting signature change are
studied. The background solution consists in a Friedmann-Lemaitre-Robertson-
Walker (FLRW) Universe filled by a constant scalar field playing the role
of a cosmological constant. It is shown that no regular solution exist
satisfying the junction conditions at the surface of change. The comparison
with similar studies in quantum cosmology is made.
\end{abstract}
\vspace{3cm}
\footnotesize${}^*$
Permanent address: Laboratoire de Gravitation  et Cosmologie Relativistes,
Universit\'e Pierre et Marie Curie,  CNRS/URA 769, Tour 22/12, Boite courier
142, 4 place Jussieu 75252 Paris  cedex 05, France.
\normalsize
\end{titlepage}
\section{Introduction}
The idea that the signature of spacetime could have been different in the very
early Universe has first appeared in association with  quantum cosmology
\cite{ref1,ref2}. Quantum cosmology is based on the ADM version of general
relativity \cite{ref3} in which spacetime is sliced into three-dimensional
spacelike hypersurfaces. The true degrees of freedom of the gravitational field
are described by the components $h_{ij}$ of the metric on these hypersurfaces
whereas the way in which they are matched remains essentially arbitrary. The
main object of quantum cosmology is the wave function of the Universe $\Psi $,
solution of the Wheeler-De Witt equation. The function $\Psi $ depends only on
$h_{ij} $ and eventually on the matter fields. For minisuperspace models,
the WKB approximation of $\Psi $ can be roughly described either by an
oscillatory function $e^{iS}$ or by the exponential $e^{-S}$ corresponding
respectively to classically allowed and classically forbidden behaviour of
the Universe. Wave functions $e^{-S}$ are similar to those used in ordinary
quantum mechanics for the study of the tunnel effect. Formally, it can be
obtained from the oscillatory wave function by a Wick rotation (or a passage
to imaginary time). In quantum cosmology, this suggests to interpret the wave
functions $\Psi \sim e^{-S}$ as describing in fact a Riemannian manifold.
Therefore, in this framework, change of signature appears as one of possible
quantum properties of spacetime. One of the main advantages of this
formulation is that it gives a solution to the initial singularity problem.
\par
Recently, it has been realized that change of signature can be also considered
in the framework of classical general relativity \cite{ref4}-\cite{ref22}.
This results from the fact that the Einstein equations do not fix the
signature which thus appears as an extra assumption added {\it a posteriori}
to the theory. Solutions very similar to the "no-boundary" solution of quantum
cosmology \cite{ref1} have been obtained \cite{ref4,ref7}. A crucial point
in the study of the classical change of signature is the choice of the
matching conditions at the surface of change. In the litterature, two
different approaches have been adopted. The first one requires the continuity
of the second fundamental form $K_{ij}$ on the surface of change $\Sigma $
\cite{ref4,ref6,ref15,ref16,ref22} whereas the second, demands in addition
that $K_{ij}$  should vanish on $\Sigma $ \cite{ref7,ref10,ref11,ref12,ref20}.
In what follows, we shall favour the second approach, in which the junction
conditions can be obtained naturally as a consequence of the distributional
parts of the field equations on $\Sigma $ \cite{ref7,ref10}. This approach
is more restrictive than the first one and the class of solutions is smaller.
Hayward \cite{ref10} has emphasized that this restrictiveness can lead to
predictions which are in agreement with the present status of the cosmological
observations. However, his conclusion is valid  only for the background
solution of Einstein equations.
\par
There is no need to emphasize the importance of theoretical models of the
cosmological perturbations \cite{ref23,ref24,ref25}. They lead to significant
results which can be compared with the observations. For example, the
presence of small perturbations can be detected by their influence upon the
angular variation of the cosmic background microwave radiation observed by
the satellite COBE \cite{ref24,ref26}. Any model attempting to describe
the primordial Universe must be able to explain these particular features
of the background radiation.
\par
The aim of this article is to address this question in the framework of the
model in which the Universe is described by a manifold displaying a classical
change of signature. More precisely, the background will consist in a FLRW
metricv and the matter will be represented by a constant scalar field
playing the role of a cosmological constant.
\par
The article is organized as follows. In the second section, we describe the
background model. For the sake of completeness, we also describe briefly the
later stages of evolution of the Universe after the inflation. The third
section is devoted to the perturbed Einstein equations with signature change.
These equations are solved in the Riemannian region, taking into account all
types of perturbations, namely the density perturbations, the rotational
perturbations and gravitational waves. In the fourth section, we compare the
solutions displaying a classical change of signature with those arising in
the framework of quantum cosmology. Conclusions are presented in the fifth
section.

\section{The background model}

The structure of the background manifold can be symbolized as follows:
\begin{equation}
\label{1}
M=M^-\cup \Sigma \cup M^+
\end{equation}
where $M^-$ ($M^+$) is a Riemannian (Lorentzian) manifold endowed with a metric
tensor whose the signature is $++++$ ($-+++$). $\Sigma $ is the
three-dimensional spacelike hypersurface where the dynamical change of
signature
occurs. On this surface, the metric is either non degenerate but discontinuous
("discontinuous proposal") or degenerate but continuous ("continuous
proposal").
It can be shown that implementing the classical change of signature with
any of these two proposals is equivalent \cite{ref20}. Here, we will adopt
the discontinuous approach. As previously emphasized in Ref. \cite{ref7,ref10},
this requires the use of distributions in the sense of Schwarz.
\par
In order to construct cosmological models, we will restrict our considerations
to FRLW manifolds in which the
spacelike sections are three-dimensional spheres. In that case, the metric
tensor can be written as:
\begin{equation}
\label{2}
g=-N(t){\rm d}t\otimes{\rm d}t+a^2(t)\biggl({\rm d}\chi \otimes {\rm
d}\chi+\sin ^2\chi ({\rm d}\theta \otimes {\rm d}\theta+\sin ^2\theta {\rm
d}\varphi \otimes {\rm d}\varphi )\biggr)
\end{equation}
The sign of the lapse function $N(t)$ fixes the signature of $g$. The matter
is described by a scalar field whose action is given by:
\begin{equation}
\label{3}
S[\phi]=\int {\rm d}^4x -\sqrt{-g}\biggl(\frac{1}{2}g^{\alpha \beta }\phi
_{,\alpha }\phi _{,\beta }+V(\phi )\biggr)
\end{equation}
Then, the system of Einstein-Klein-Gordon equations takes on the form
(a dot means a derivative with respect to $t$):
\begin{eqnarray}
\label{4}
& &\frac{\dot{a}^2}{a^2}+\frac{N}{a^2} = \frac{1}{6}\biggl( \frac{\dot{\phi
}^2}{2}+NV(\phi )\biggr) \\
\label{5}
& &\frac{\ddot{a}}{a}-\frac{\dot{N}\dot{a}}{2aN} = -\frac{1}{6}\biggl(\dot{\phi
}^2-NV(\phi )\biggr) \\
\label{6}
& &\ddot{\phi }-\frac{\dot{N}\dot{\phi }}{2N}+3\frac{\dot{a}}{a}\dot{\phi
}+N\frac{{\rm d}V}{{\rm d}\phi}=0
\end{eqnarray}
In the discontinuous proposal, the lapse function $N(t)$ is a distribution
equal to:
\begin{equation}
\label{7}
N(t)=2Y(t)-1=\epsilon
\end{equation}
where $Y(t)$ denotes the Heaviside distribution. $\epsilon $ is equal to $-1$
in the Riemannian region and to $+1$ in the Lorentzian one. The time derivative
of the lapse distribution is equal to $2\delta (\Sigma )$ where $\delta (\Sigma
)$ is the Dirac distribution on the surface of change $\Sigma $. Then, the
Einstein-Klein-Gordon system of equations splits up into the regular part:
\begin{eqnarray}
\label{8}
& &\frac{\dot{a}^2}{a^2}+\frac{\epsilon}{a^2} = \frac{1}{6}\biggl(
\frac{\dot{\phi }^2}{2}+\epsilon V(\phi )\biggr) \\
\label{9}
& &\frac{\ddot{a}}{a} = -\frac{1}{6}\biggl(\dot{\phi }^2-\epsilon V(\phi
)\biggr)
\\
\label{10}
& &\ddot{\phi }+3\frac{\dot{a}}{a}\dot{\phi}+\epsilon \frac{{\rm d}V}{{\rm
d}\phi}=0
\end{eqnarray}
and, since we should treat the distributional part separately
\cite{ref7,ref10}, to
the singular part:
\begin{eqnarray}
\label{11}
\dot{a}\delta (\Sigma ) &=& 0 \\
\label{12}
\dot{\phi }\delta (\Sigma ) &=& 0
\end{eqnarray}
which produces the following junction conditions at the surface of change:
$\dot{a}=0$ and $\dot{\phi }=0$. In what follows in this article, we shall
consider only the case where $\dot{\phi }=0$ everywhere. This implies that
on the background level, the scalar field plays the role of the cosmological
constant. Then, the solutions of field equations satisfying the matching
conditions read :
\begin{eqnarray}
\label{13}
a(t) &=& \frac{1}{H}\cos Ht \qquad -\frac{\pi }{2H}\leq t\leq 0 \\
\label{14}
a(t) &=& \frac{1}{H}\cosh Ht \qquad t \geq 0
\end{eqnarray}
where $H$ is defined as $\sqrt{\frac{V}{6} }$. In the framework of classical
general relativity this solution was first discovered by Ellis et al.
\cite{ref4} and Hayward \cite{ref7}. In the Riemannian region, the manifold
is the sphere $S^4$ and in the Lorentzian region the Universe undergoes an
inflationary phase described by a half of the De Sitter spacetime. We have
thus constructed a cosmological model without the initial singularity.
However, it is important to note that it does not violate the Hawking-Penrose
theorem on singularities \cite{ref27} since we have given up one of the
fundamental assumptions necessary to prove this theorem, namely the causal
structure of spacetime. It is also worth reminding that $M$ can be obtained
from the entire De Sitter spacetime by performing a Wick rotation in the lower
part of the hyperboloid. As the De Sitter spacetime is also a singularity-free
manifold, this shows that the classical change of signature cannot be used in
order to avoid any singularity. It simply enables us to construct
singularity-free cosmological models which are, in a certain sense, "finite
in time". This result is a particular case of a theorem proved in Ref.
\cite{ref20}.
\par
The solutions (\ref{13})-(\ref{14}) are very similar to those obtained in
quantum cosmology. The corresponding model is a one-dimensional minisuperspace
for which the Wheeler-De Witt equation can be expressed in the following form
\cite{ref1,ref2}:
\begin{equation}
\label{15}
\biggl(\frac{1}{4}\pi _a^2-a^2(a^2\frac{V(\phi )}{6}-1)\biggr)\Psi (a, \phi)=0
\end{equation}
The minisuperspace can be divided into two parts: one for which the potential
$U(\phi )\equiv a^2(a^2\frac{V(\phi )}{6}-1)$ is negative, i.e for which
$a<\frac{1}{H}$ and one for which $U(\phi )$ is positive, i.e for which
$a>\frac{1}{H}$. The corresponding WKB solution of equation (\ref{15}) is
either an exponential or an oscillatory function, thus showing that the
Riemannian region is classically forbidden in quantum cosmology. It is not
clear if we should consider the model with the classical change of signature
as a limit of a quantum solution when the Planck constant tends to zero, or if
we should take the classical model into account seriously and then try to
quantize it as it has been proposed in Ref. \cite{ref19}. The matching
conditions can be recovered by using the fact that at the surface of change
$K_{ij}$ must vanish. In such case, it is equivalent to set $\dot{a}=0$.
\par
We can also construct a complete cosmological scenario which consists
in radiation-dominated and matter-dominated eras following the inflationary
stage. In the region $t\geq 0$, we use for our convenience the conformal
time $\eta $ defined by ${\rm d}\eta = {\rm d}t/a$. The integration of the
previous equation leads to the following relationship: $\eta =2\arctan (e^{
Ht})$, $0<\eta <\pi$. In terms of $\eta $-time,
the scale factor in the inflationary region can be written as:
\begin{equation}
\label{16}
a(\eta ) = \frac{1}{H\sin \eta} \\
\end{equation}
We assume that for $\eta =\eta _1$, the inflation stops and that the Universe
enters a radiation-dominated era characterized by the equation of state
$p=\rho /3$, for which the behaviour of the scale factor is given by:
\begin{equation}
\label{17}
a(\eta )=a_r\sin (\eta -\eta _r)
\end{equation}
The constants $a_r$ and $\eta _r$ are chosen in such a way that the function
$a(\eta )$ is $C^1$ when $\eta =\eta _1$. This provides the following equations
for $a_r$ and $\eta _r$:
\begin{eqnarray}
\label{18}
\eta _r &=&  2\eta _1 \\
\label{19}
a_r &=& -\frac{1}{H\sin ^2\eta _1}
\end{eqnarray}
After the radiation-dominated era, the matter-dominated era begins at
$\eta =\eta _2$. The equation of state is now $p=0 $ and the solution for
the scale factor is:
\begin{equation}
\label{20}
a(\eta )=\frac{a_m}{2}\biggl(1-\cos(\eta -\eta _m)\biggr)
\end{equation}
The requirement that $a(\eta )$ and $a'(\eta )$ must be continuous at
$\eta =\eta_2$ is expressed by the following two equations:
\begin{eqnarray}
\label{21}
\eta _m &=& -\eta _2+2\eta _r \\
\label{22}
a_m &=& \frac{a_r}{\sin (\eta _2-\eta _r)}
\end{eqnarray}
The matter era represents the present state of our Universe. We have now at
our disposal a complete description of the evolution of the Universe.
\section{The perturbed model for the Riemannian region}
\subsection{General equations}
This section is devoted to the study of perturbations around the Riemannian
model described before. We shall assume that in the Riemanian region, the
geometry could have fluctuated around the sphere $S^4$. These "deformations"
(probably a more appropriate term than "perturbations" since, in this region,
there is no notion of evolution) could constitute a possible mechanism
explaining the origin of the perturbations in the usual (Lorentzian) Universe.
Another interesting point is the following: the ordinary theory of cosmological
perturbations enables to compute the evolution of the perturbed metric but does
not fix the initial conditions. To specify them, the principles of quantum
mechanics are usually evoked \cite{ref24}; the initial state of the
perturbations is taken to be the vacuum state. In the model of classical
change of signature, we can also hope to say something about the initial data
for the perturbations. Indeed, we will have to solve the equations of motion
which are second-order differential equations. The initial data for an ordinary
second-order differential equation are $f(t_0)$ and $f'(t_0)$ for a given
initial time $t=t_0$. As emphasized in Ref. \cite{ref10}, this is no longer the
case for the singular equations which arise in the framework of the classical
change of signature. The initial data are now $f(t_0)$ with the requirement
that $f'(t_0)=0$ ($t_0$ is the value of $t$ for which the change of signature
occurs). In a certain sense, the change of signature fixes the initial
conditions for the perturbations (however, we still have the freedom to choose
$f(t_0)$). We will study later the consequences of this fact.
\par
Let us consider a small perturbation $\delta g_{\mu \nu }=h_{\mu \nu }$ of the
background metric (\ref{2}). The physical metric can be written as:
\begin{equation}
\label{23}
g_{\mu \nu}=g_{\mu \nu}^{(0)}+h_{\mu \nu}
\end{equation}
It is well known that $h_{\mu \nu}$ contains unphysical degrees of freedom.
Therefore, we will perform the computation using the synchronous gauge, that
is, by requiring $h_{0\mu }=0$. This choice has two main advantages. The first
one is a technical reason: with this gauge, the calculations are easier;
the second one is related to the interpretation of the solutions. Had we not
have chosen the synchronous gauge, the frontier between the Riemannian  and
Lorentzian regions would have been blurred due to the presence of a term
$h_{00}$, possibly positive or negative. With this gauge, the frontier
still remains distinct. The components of the perturbed Ricci tensor will be
given in terms of $h^i{}_j=g^{(0)ik}h_{kj}$ and $h$ will denote the trace of
the perturbed metric, $h^i{}_i=h$. The symbol "$\ \ \tilde{} \ \ $" will refer
to the three-dimensional metric $\tilde{g}_{ij}$ related to $g$ by the
equation:
\begin{equation}
\label{24}
g=-N(t){\rm d}t\otimes {\rm
d}t+a^2(t)\tilde{g}_{ij}{\rm d}x^i\otimes {\rm d}x^j
\end{equation}
The components of the perturbed Ricci tensor can be written as:
\begin{eqnarray}
\label{25}
\delta R^0{}_0 &=&
\frac{1}{2N}(\ddot{h}+2\frac{\dot{a}}{a}\dot{h})-\frac{\dot{N}}{2N^2}\dot{h} \\
\label{26}
\delta R^0{}_i &=& \frac{1}{2N}\frac{{\rm \partial }}{{\rm \partial
}t}(\tilde{\nabla }_ih-\tilde{\nabla }_kh^k{}_i) \\
\label{27}
\delta R^i{}_j &=&
\frac{1}{2N}\ddot{h}^i{}_j-\frac{3}{2N}\frac{\dot{a}}{a}\dot{h}
^i{}_j+\frac{1}{2N}\frac{\dot{a}}{a} \dot{h}^k{}_k\delta
^i{}_j-\frac{2}{a^2}h^i{}_j-\frac{\dot{N}}{4N^2}\dot{h}^i{}_j  \nonumber \\
& &-\frac{1}{2a^2}\biggl(\tilde{\nabla }_j\tilde{\nabla }^ih^k{}_k
-\tilde{\nabla }^l\tilde{\nabla }_jh^i{}_l
-\tilde{\nabla }_l\tilde{\nabla }^ih^l{}_j
+\tilde{\nabla }^k\tilde{\nabla }_kh^i{}_j \biggr)
\end{eqnarray}
It is interesting to note the presence of additional terms due to the lapse
function. These terms will provide the junction conditions for the perturbed
metric. According to Lifshitz and Khalatnikov \cite{ref23}, we can classified
the perturbations as three types: scalar, vector and tensor ones. This
classification is based on a theorem which states that any tensor of rank
two can be decomposed as:
\begin{equation}
\label{28}
h^i{}_j=h^i{}_j{}^{(S)}+h^i{}_j{}^{(V)}+h^i{}_j{}^{(TT)}
\end{equation}
where $\tilde{\nabla }_ih^i{}_j{}^{(V)}=0$, $h^k{}_k{}^{(TT)}=0$ and
$\tilde{\nabla }_ih^i{}_j{}^{(TT)}=0$. The symbol "$(TT)$" means transverse and
trace-free. We can treat each type separately. The tensorial part
$h^i{}_j{}^{(TT)}$ represents primordial gravitational waves whereas
$h^i{}_j{}^{(S)}$ and  $h^i{}_j{}^{(V)}$ represent respectively density
and rotational perturbations, i.e. the perturbations accompanying fluctuations
of the matter filling the Universe.
\subsection{Density and rotational perturbations}
Let us consider first the case of the density perturbations. They are
constructed using the eigenfunction $Q(\chi, \theta, \varphi )$ of the
three-dimensional Laplacian \cite{ref23}:
\begin{equation}
\label{29}
\tilde{\nabla }^k\tilde{\nabla }_kQ(\chi, \theta, \varphi )=
-(n^2-k)Q(\chi, \theta, \varphi )
\end{equation}
where $n$ is an integer greater or equal to one. Following Ref. \cite{ref23},
we can define the following tensors:
\begin{eqnarray}
\label{30}
Q^i{}_j &=& \frac{Q}{3}\delta ^i{}_j \\
\label{31}
P^i{}_j &=& \frac{\tilde{\nabla }^i\tilde{\nabla }_jQ}{n^2-k}+Q^i{}_j
\end{eqnarray}
and express the scalar part of the perturbed metric $h^i{}_j{}^{(S)}$ as:
\begin{equation}
\label{32}
h^i{}_j{}^{(S)}=\lambda (\eta )P^i{}_j+\mu (\eta )Q^i{}_j
\end{equation}
In this section, for convenience, we do not write the indices and the sums. In
fact $Q$ has to be understood as $Q^n_{lm}$ and $\lambda (\eta )$, $\mu (\eta
)$ as $\lambda _{nlm}(\eta )$ and $\mu _{nlm}(\eta )$. Then, it turns out that
the components of $\delta R^{\mu }{}_{\nu }{}^{(S)}$ are given by:
\begin{eqnarray}
\label{33}
\delta R^0{}_0{}^{(S)} &=& \frac{1}{2N}(\ddot{\mu }+2\frac{\dot{a}}{a}\dot{\mu
}-\frac{\dot{N}}{N}\dot{\mu })Q \\
\label{34}
\delta R^0{}_i{}^{(S)} &=& \frac{1}{3N}\biggl((n^2-k)\dot{\mu
}+(n^2-4k)\dot{\lambda }\biggr)\frac{\tilde{\nabla }_iQ}{n^2-k}  \\
\label{35}
\delta R^i{}_j{}^{(S)} &=& \biggl(\frac{\ddot{\mu
}}{2N}+\frac{3\dot{a}}{Na}\dot{\mu }-\frac{\dot{N}\dot{\mu
}}{4N^2}+2\frac{n^2-4k}{3a^2}(\lambda +\mu)\biggr) Q^i{}_j \nonumber \\
& & +\biggl(\frac{\ddot{\lambda }}{2N}+\frac{3\dot{a}}{2Na}\dot{\lambda }
-\frac{\dot{N}\dot{\lambda }}{4N^2}-\frac{n^2-k}{6a^2}(\lambda
+\mu)\biggr)P^i{}_j
\end{eqnarray}
The components of the perturbed source tensor, defined as $S_{\mu \nu}=T_{\mu
\nu}-\frac{T}{2}g_{\mu \nu}$, are:
\begin{eqnarray}
\label{36}
\delta S^0{}_0 &=& \frac{{\rm d}V}{{\rm d}\varphi }\delta \varphi \\
\label{37}
\delta S^0{}_i &=& 0 \\
\label{38}
\delta S^i{}_j &=& \frac{{\rm d}V}{{\rm d}\varphi }\delta \varphi \delta ^i{}_j
\end{eqnarray}
If the function $\delta \varphi $ is decomposed according to $\delta \varphi
=f(t)Q(\chi, \theta, \varphi )$, the regular part of the perturbed Einstein
equations takes on the form (in fact $\epsilon =-1$ since we study the
perturbations in the Riemanian region):
\begin{eqnarray}
\label{39}
\ddot{\mu }+2\frac{\dot{a}}{a}\dot{\mu } &=& 2\epsilon
\frac{{\rm d}V}{{\rm d}\varphi }f \\
\label{40}
(n^2-k)\dot{\mu }+(n^2-4k)\dot{\lambda } &=& 0 \\
\label{41}
\frac{\ddot{\lambda }}{2}+\frac{3\dot{a}}{2a}\dot{\lambda
}-\epsilon \frac{n^2-k}{6a^2}(\lambda +\mu) &=& 0 \\
\label{42}
\frac{\ddot{\mu }}{2}+\frac{3\dot{a}}{a}\dot{\mu
}+2\epsilon \frac{n^2-4k}{3a^2}(\lambda +\mu) &=& 3\epsilon
\frac{{\rm d}V}{{\rm d}\varphi }f
\end{eqnarray}
and the distributional part reduces to:
\begin{equation}
\label{43}
\dot{\mu }\delta (\Sigma )=\dot{\lambda }\delta (\Sigma )=0
\end{equation}
Therefore, the junction conditions for the density perturbations are $\dot{\mu
}=\dot{\lambda }=0$ on the surface of change. From equations
(\ref{39})-(\ref{42}), one can compute the combination
$(n^2-4k)(\ref{41})+\frac{n^2-k}{4}(\ref{42})+\frac{3}{4}(n^2-k)(\ref{39})$.
One obtains:
\begin{equation}
\label{44}
\frac{1}{2}\biggl((n^2-k)\ddot{\mu }+(n^2-4k)\ddot{\lambda
}\biggr)+\frac{3\dot{a}}{2a}\biggl((n^2-k)\dot{\mu }+(n^2-4k)\dot{\lambda
}\biggr)=\frac{3}{2}\epsilon \frac{{\rm d}V}{{\rm d}\varphi }f
\end{equation}
A comparison with equation (\ref{40}) shows that $V_{,\varphi }=0$. Then, the
integration of equation (\ref{39}) provides the following solution for $\mu $:
\begin{equation}
\label{45}
\mu =\alpha \int \frac{{\rm d}t}{a^2}+\beta
\end{equation}
where $\alpha $ and $\beta $ are arbitrary constants. Taking into account the
matching conditions, we see that $\alpha $ must vanish. Therefore, the
solutions of the Einstein equations (\ref{39})-(\ref{42}) satisfying the
junction conditions are $\mu =-\lambda=\beta $. However, the relationship
$h_{0\mu }=0$ does not fix the gauge completely  \cite{ref28} which is
preserved under the following coordinate transformations:
\begin{eqnarray}
\label{46}
t &\rightarrow & t \\
\label{47}
x_i &\rightarrow & x_i+a^2f_i(x^k)
\end{eqnarray}
where $f_i(x^k)$ is an arbitrary function of the spacelike coordinates. Then,
the scalar part of the perturbed metric transforms according to
$h^i{}_j{}^{(S)} \rightarrow h^i{}_j{}^{(S)}+
2A\tilde{\nabla }^i\tilde{\nabla }_jQ$. This implies $\lambda \rightarrow
\lambda +2(n^2-k)A$ and $\mu \rightarrow  \mu +2(n^2-k)A$. Therefore, with a
proper choice of $A$ ($A=\frac{\beta }{2(n^2-k)}$) we conclude that the
density perturbations do vanish identically.
\par
Let us turn now to the case of rotational perturbations. Their construction
implies the transverse eigenvector $S_i$ ($\tilde{\nabla }^iS_i=0$) of the
three-dimensional Laplacian \cite{ref23}:
\begin{equation}
\label{48}
\tilde{\nabla }^k\tilde{\nabla }_kS_i(\chi, \theta, \varphi )=
-(n^2-2k)S_i(\chi, \theta, \varphi )
\end{equation}
where $n$ is an integer greater or equal to $2$. If we introduce the tensor
$S_{ij}=-\frac{1}{2}(\tilde{\nabla }_iS_j+\tilde{\nabla }_jS_i)$ and define
$h^i{}_j{}^{(V)}$ as  $h^i{}_j{}^{(V)}=\sigma S^i{}_j$, then it turns out
that the components of the perturbed Ricci tensor can be written as:
\begin{eqnarray}
\label{49}
\delta R^0{}_0{}^{(V)} &=& 0\\
\label{50}
\delta R^0{}_i{}^{(V)} &=& \frac{\dot{\sigma }}{2N}(n^2-4k)S_i \\
\label{51}
\delta R^i{}_j{}^{(V)} &=& (\frac{\ddot{\sigma
}}{2N}+\frac{3\dot{a}}{2Na}\dot{\sigma }-\frac{\dot{N}}{4N^2}\dot{\sigma
})S^i_j
\end{eqnarray}
The perturbed stress-energy tensor does not contain any rotational terms,
since a scalar field cannot support rotational oscillations. Therefore, the
regular part of the Einstein equations reduces to:
\begin{eqnarray}
\label{52}
\frac{\dot{\sigma }}{2}(n^2-4k) &=& 0 \\
\label{53}
\ddot{\sigma }+3\frac{\dot{a}}{a}\dot{\sigma } &=& 0
\end{eqnarray}
whereas the distributional part is:
\begin{equation}
\label{54}
\dot{\sigma }\delta (\Sigma )=0
\end{equation}
The junction condition for the rotational perturbation is $\dot{\sigma }=0$ at
the surface of change. It is straightforward to integrate this equation; the
solution is simply a constant:
\begin{equation}
\label{55}
\sigma (t)=C
\end{equation}
Under the residual gauge (\ref{46})-(\ref{47}), $\sigma $ transforms according
to $\sigma \rightarrow \sigma +B$ where $B$ is an arbitrary constant. The
choice $B=-C$ shows that the rotational perturbations can not exist either.
\par
So, we have reached the conclusion that the perturbations associated with
fluctuations within the matter do not survive in the Riemannian region $S^4$.
This is not surprising and it is not a special feature of the solutions with
change of signature. Indeed, this fact is already known for the De Sitter
spacetime \cite{ref29}. Here, we simply recover the same result because $S^4$
is nothing else but the De Sitter spacetime with imaginary time.
\subsection{Gravitational waves}
Contrary to the density and rotational perturbations, tensorial perturbations
represent fluctuations of the geometry only. Therefore, it is for this type
of perturbations that we can expect to obtain informations concerning the
primordial Riemannian region. The tensor $h^i{}_j{}^{(TT)}$ can be expressed
by means of tensor spherical harmonics on $S^3$. Tensor spherical harmonics
are the eigentensors of the Laplacian on the three-dimensional sphere and are
defined by the equation \cite{ref23}:
\begin{equation}
\label{56}
\tilde{\nabla }^k\tilde{\nabla }_kG_{ij}(\chi,\theta, \varphi )=
-(n^2-3k)G_{ij}(\chi, \theta, \varphi )
\end{equation}
with $\tilde {\nabla }^iG_{ij}=0$, $\tilde{g}^{ij}G_{ij}=0$. $n$ is an
integer greater than three. The explicit form of the eigentensors can be found
in Ref. \cite{ref30}. They are normalized and obey to the relationship:
\begin{equation}
\label{57}
\int {\rm d}^3x\sqrt{\tilde{g}}(G_{ij})^n_{lm}(G^{ij})^{n'}_{l'm'}=\delta
^{nn'}\delta _{ll'}\delta _{mm'}
\end{equation}
As a consequence, we can develop $h^i{}_j{}^{(TT)}$ in the basis of the
eigentensors $(G^{}_j)^n_{lm}$:
\begin{equation}
\label{58}
h^i{}_j{}^{(TT)}(t, \chi, \theta, \varphi )=\sum ^{\infty }_{n=3}\sum
^{n-1}_{l=2}\sum ^{l}_{m=-l} \nu _{nlm}(t)(G^i{}_j)^n_{lm}(\chi, \theta,
\varphi )
\end{equation}
Putting this expression in equation (\ref{25})-(\ref{27}) enables us to compute
the components of the perturbed Ricci tensor:
\begin{eqnarray}
\label{59}
\delta R^0{}_0{}^{(TT)} &=& \delta R^0{}_i{}^{(TT)}=0 \\
\label{60}
\delta R^i{}_j{}^{(TT)} &=& (\frac{\ddot{\nu _N
}}{2N}+\frac{3\dot{a}}{2Na}\dot{\nu _N
}+\frac{n^2-1}{2a^2}\nu _N-\frac{\dot{N}}{4N^2}\dot{\nu _N})G^i{}_j
\end{eqnarray}
where the indices and the sums have been omitted for simplicity and the index
$N$ denotes the whole set of indices $N\equiv (n,l,m)$. In what follows,
we will consider each mode $(n,l,m)$ separately. Since the perturbed  source
tensor does not contain any tensorial part, the perturbed Einstein  equations
for the gravitational waves reduce themselves to:
\begin{equation}
\label{61}
\delta R^{\mu }{}_{\nu }{}^{(TT)}=0
\end{equation}
Therefore, the time evolution of the amplitude of each mode is determined by
the equations:
\begin{eqnarray}
\label{62}
\ddot{\nu _N}+3\frac{\dot{a}}{a}\dot{\nu _N}+\epsilon \frac{n^2-1}{a^2}\nu _N=0
\\
\label{63}
\dot{\nu _N}\delta (\Sigma )=0
\end{eqnarray}
The distributional part (\ref{63}) of Einstein's equations provides the
junction condition for each mode:
\begin{equation}
\label{64}
\frac{{\rm d}\nu _N}{{\rm d}t}=0
\end{equation}
The next step is to solve the equation (\ref{62}) in the Riemannian region,
that is to say for $a(t)=\frac{1}{H}\cos Ht$ and $\epsilon =-1$. In that case,
one obtains:
\begin{equation}
\label{65}
\frac{{\rm d}^2\nu _N}{{\rm d}t^2}-3H\frac{\sin Ht}{\cos Ht}\frac{{\rm d}\nu _N
}{{\rm d}t}-H^2\frac{n^2-1}{\cos ^2Ht}\nu _N=0
\end{equation}
Let us study the features of this equation in more detail. For convenience,
let us introduce the coordinate $\tau $ defined by $\tau =Ht+\frac{\pi}{2}$
such that the south pole is now given by $\tau =0$. Near the south pole, for
small values of $\tau $, the solution is $\nu _N\sim A\tau ^{n-1}+B\tau
^{-(n+1)}$ revealing that equation (\ref{62}) possesses one regular and one
divergent solution. However, since the system of coordinates is not well
defined at the south pole, this does not imply anything in what concerns the
real behaviour of the deformations at this point. To deal with this problem,
let us introduce a new system of coordinates. Each section of the sphere
$S^4$ is a sphere $S^3$ which can be embedded into the four-dimensional
Euclidean space $R^4$. A point of $S^3$ can be located with the coordinate
$\bar{x}^A$, $A=1,...,4$, satisfying the constraint $\bar{x}^A\bar{x}_A=r^2$
where $r$ is the radius of $S^3$ (see figure 1). Then, the following formulae
hold  \cite{ref23,ref31}:
\begin{equation}
\label{66}
G^{(n)}_{ij}(\chi, \theta, \varphi ){\rm d}x^i{\rm
d}x^j=\frac{1}{r^{n-1}}T^{(n)}_{AC_1BC_2...C_{n-1}}\bar{x}^{C_1}...\bar{x}^
{C_{n-1}}{\rm d}\bar{x}^A{\rm d}\bar{x}^B
\end{equation}
where $x^i=(\chi, \theta, \varphi )$ and $T^{(n)}_{AC_1BC_2...C_{n+1}}$ is a
constant tensor of rank $n+1$ with the following properties:
\cite{ref23,ref31}.
It is antisymetric with respect to the pairs of indices $AC_1$ and $BC_2$,
symmetric in the $C_n$ indices ($n\geq 3$), trace-free in any pair of indices
and gives zero if we take the cyclic sum over any three different indices.
$T^{(n)}_{AC_1BC_2...C_{n+1}}\bar{x}^{c_1}...\bar{x}^{c_{n-1}}$ is an
homogeneous polynomial of order $n-1$. The unit sphere $S^4$ can be projected
perpendicularly onto the four-dimensional plan $P_S$. This corresponds to
the following choice for the coordinate $\bar{x}^A$:
\begin{equation}
\label{67}
\bar{x}^A=rf(\chi, \theta, \varphi)=\sin \tau f^a(\chi, \theta, \varphi )
\end{equation}
The resulting background metric is now regular at the south pole. Note that
this new metric is no longer diagonal. To preserve the gauge, we could have
used the stereographic projection ($M$ becomes $M''$ in figure 1):
\begin{equation}
\label{68}
\bar{x}^A=2\tan \frac{\tau }{2}f^A(\chi, \theta, \varphi )
\end{equation}
However, as we are interested in the behaviour of the solution when $\tau $
tends to zero, the two previous changes of coordinates are equivalent for our
purpose and we shall work with the first one. Under the transformation
(\ref{67}), the metric becomes:
\begin{eqnarray}
\label{69}
 g &=& g^{(0)}+\sum^{\infty }_{n\geq 3}\nu _n(\tau
)T^{(n)}_{AC_1BC_2...C_{n-1}}
f^{C_1}...f^{C_{n-1}}{\rm d}\bar{x}^A{\rm d}\bar{x}^B
\end{eqnarray}
Equation (\ref{69}) demonstrates explicitely that the "true" behaviour of the
deformations is determined by the behaviour of the functions $\nu _N$. As a
consequence, one solution for the deformations is indeed regular at the
south pole whereas the other one actually diverges.
\par
Let us now study the exact solutions of equation (\ref{62}). The change of
variable $x=\sin Ht$ and the change of function $\nu =\sqrt{1-x^2}g$ reduce
our equation to:
\begin{equation}
\label{70}
(1-x^2)g''-2xg'+(2-\frac{n^2}{1-x^2})g=0
\end{equation}
which is a Legendre differential equation. The general solution is given in
terms of the Legendre functions $P^{-n}_1$ and $Q^n_1$
\cite{ref32,ref33,ref34}:
\begin{equation}
\label{71}
\nu _{nlm}(t)=\frac{A^R_{nlm}}{\cos Ht}P^{-n}_1(\sin Ht)+\frac{B^R_{nlm}}{\cos
Ht}Q^n_1(\sin Ht)
\end{equation}
where $A^R_{nlm}$ and $B^R_{nlm}$ are two arbitrary constants. We can also
express the solution in a representation which is more convenient for the
study of matching conditions:
\begin{eqnarray}
\label{72}
\nu _{3lm}(t) &=& C^R_{3lm}(-8\tan ^4Ht -12\tan ^2Ht -3)+D^R_{3lm}\frac{\sin
Ht}{\cos ^4Ht} \\
\label{73}
\nu _{4lm}(t) &=& C^R_{4lm}(-24\tan ^5Ht -40\tan ^3Ht -15\tan Ht) \nonumber \\
& & +D^R_{4lm}\frac{1+5\sin ^2Ht}{\cos ^5Ht}
\end{eqnarray}
where $C^R_{nlm}$ and $D^R_{nlm}$ are arbitrary constants. For higher values
of $n$, the general form is preserved, i.e a polynomial of order $n+1$
in $\tan Ht$ for the first branch and a polynomial of order $n-2$ in $\sin Ht$
divided by $\cos ^{n+1}Ht$ for the second branch. For each value of $(n,l,m)$,
both branches blow up at $t=-\pi /2H$. However, according to the previous
discussion, a specific choice of $C^R_{nlm}$ and $D^R_{nlm}$ always enables us
to construct a solution regular at the south pole. For example, for $n=3$, the
choice is $-8C^R_{3lm}=D^R_{3lm}$. The next step is to take into account the
matching conditions (\ref{63}). They require  $C^R_{2qlm}=D^R_{2p+1lm}=0$ with
$q\geq 2$ and $p\geq 1$. In other words, only one branch (not always the same
according to if $n$ is a even or odd number) can cross the surface of change.
Therefore, we have reached the conclusion that the requirement of regularity
of the solution in the Riemannian region and the junction conditions are
incompatible: every solution satisfying $\frac{{\rm d}\nu _N}{{\rm d}t}=0$ is
divergent at the south pole, except the trivial function $\nu _N=0$. Note
that it is known that the sphere $S^4$ is an isolated solution of the
Riemannian Einstein equations \cite{ref34.5}. In this paper, we recover this
result using the formalism of cosmological pertubations. Before
discussing the consequences of this result for the classical change of
signature, we are going to study the equivalent problem in the framework of
quantum cosmology.
\section{Quantum cosmology and classical change of signature}
In this section, we follow the treatment given by Halliwell and Hartle
\cite{ref35}. The basic formulae of the path integral formulation of quantum
gravity is:
\begin{equation}
\label{74}
\Psi[\bar{h}_{ij},\bar{\Phi },B]=\sum _{M}\int _{{\cal C}} {\cal D}g_{\mu \nu}
{\cal D}\Phi e^{-I[g_{\mu \nu},\Phi]}
\end{equation}
where the sum is taken over all manifolds $M$ having $B$ as part of their
boundary and over all metrics $g_{\mu \nu}$ and matter fields $\Psi $ which
induce $\bar{h}_{ij}$ and $\bar{\Phi }$ on $B$. $I$ denotes the Euclidean
Einstein-Hilbert action. Computing the wave function $\Psi $ is a difficult
task and this question has been discussed extensively in the litterature. The
action $I$ is unbounded from below for real metric. Then, in order to give a
meaning to $\Psi $, we must perform the integration over complex metrics
\cite{ref35,ref36}. This can be done explicitly for minisuperspace models. In
that case, it has been shown by Halliwell \cite{ref37} that the propagator
between fixed three-geometry, in the gauge $\dot{N}=0$, is given by:
\begin{equation}
\label{75}
G(q''^{\alpha }|q'^{\alpha })=\int {\rm d}N \int {\cal D}q^{\alpha
}e^{-I[q^{\alpha }(\tau ),N(\tau )]}
\end{equation}
where $q^{\alpha }$ are the coordinates in the minisuperspace. $G$ is computed
over paths $q^{\alpha }(\tau )$ satisfying $q^{\alpha }(\tau ')=q'^{\alpha }$
and $q^{\alpha }(\tau '')=q''^{\alpha }$. The path integral (\ref{75}) will be
dominated by the saddle point $(N, q^{\alpha })$, that is to say by complex
configurations for which $\frac{{\rm \delta }I}{{\rm \delta }q}=0$, $\frac{{\rm
\partial }I}{{\rm \partial }N}=0$. Solving these equations provides a solution
for $q^{\alpha }$ and $N$. In general, $N$ and $q^{\alpha }$ are complex
numbers. Now, we are interested in computing $I$ for the saddle point. For
minisuperspace models, $I$ is given by:
\begin{equation}
\label{76}
I=\int _{\tau '}^{\tau ''}{\rm d}\tau N\biggl(\frac{1}{2N^2}f_{\alpha \beta }
\dot{q}^{\alpha }\dot{q}^{\beta }+U(q^{\alpha })\biggr)
\end{equation}
The solution $q^{\alpha }(\tau )$ has the property that it depends only on
$N\tau $. Then, in term of the complex variable $T=N(\tau -\tau ')$, $I$ can be
written as:
\begin{equation}
\label{77}
I=\int _{{\cal C}}{\rm d}T\biggl(\frac{1}{2}f_{\alpha \beta }\frac{{\rm
d}q^{\alpha }}{{\rm d}T}\frac{{\rm d}q^{\beta }}{{\rm d}T}+U(q(T))\biggr)
\end{equation}
where ${\cal C}$ is the contour indicated on the figure $2$, namely a straight
line from $0$ to $\bar{T}=N(\tau ''-\tau ')$. This contour can be deformed
in such a way that:
\begin{equation}
\label{78}
I=\int _{{\cal C}_1}{\rm d}T\biggl(\frac{1}{2}f_{\alpha \beta }\frac{{\rm
d}q^{\alpha }}{{\rm d}T}\frac{{\rm d}q^{\beta }}{{\rm d}T}+U(q(T))\biggr)
+\int _{{\cal C}_2}{\rm d}T\biggl(\frac{1}{2}f_{\alpha \beta }\frac{{\rm
d}q^{\alpha }}{{\rm d}T}\frac{{\rm d}q^{\beta }}{{\rm d}T}+U(q(T))\biggr)
\end{equation}
Where the contours ${\cal C}_1$ and ${\cal C}_2$ are defined as in the figure
$2$.
Along ${\cal C}_1$, $T$ is real and the corresponding solution is Riemannian.
Along ${\cal C}_2$, $T$ can be chosen in such a way that $T=\Re (\bar{T})+it\Im
(\bar{T})$, $0\leq t\leq1$, showing that the action will be purely imaginary if
$q^{\alpha }(\tau )$ is real. $q^{\alpha }(\tau )$ is real if $\dot{q}^{\alpha
}(\Re (\bar{T}))=0$ or, equivatentely, if $K_{ij}=0$. In that case, the complex
solution which dominated $G(q''^{\alpha }|q'^{\alpha })$ can be viewed as a
combination of a Riemannian and a Lorentzian manifolds. However, the point is
that in general, $G(q''^{\alpha }|q'^{\alpha })$ is dominated by an
intrinsically complex solution where all the $\dot{q}^{\alpha }$ do not vanish
for $T=\Re (\bar{T})$.
\par
The relationship with the classical change of signature is now clear. Solutions
which are a combination of a Riemannian and a Lorentzian manifolds ("tunneling
solutions") can be obtained classically by relaxing the assumption that the
signature of the metric is always hyperbolic whereas the other ones cannot be
conceived as a classical manifold with a change of signature. Our background
model belongs to the first category whereas the perturbed solution belongs to
the second. This is the reason why we found that the only solution satisfying
the regularity condition and the matching conditions was $\nu _N=0$. The
junction conditions for the first category of solutions is $K_{ij}=0$ because
$K_{ij}$ is purely real in the Riemannian region and purely imaginary in the
Lorentzian one \cite{ref35}. For the complex solutions (second category)
$K_{ij}$ has just to be continuous.
\section{Discussion and conclusion}
We are now in position to answer our main question with regard to the possible
observable consequences of the primordial change of signature. The previous
result shows that if we try to describe the primordial Universe as a FLRW
signature-changing manifold with a cosmological constant (i.e. half of the
sphere $S^4$ joined to half of the De Sitter spacetime), then it implies the
absence of primordial gravitational waves. This result is a consequence
of the fact that the sphere $S^4$ is an isolated solution of the
Riemannian Einstein equations \cite{ref34.5}. Primordial gravitational
waves are present in the context of quantum
cosmology. The reason is that the wave function of the perturbations is
dominated by a configuration which cannot be thought as a combination of a
Riemannian and a Lorentzian metric contrary to the wave function of the
background. Clearly, if we consider seriously the classical change of signature
as a possible model for describing the primordial Universe, that is to
say if we assume that not only the background solution but also the
perturbations can be represented by a signature-changing manifold (a priori,
it seems to be the most logical attitude: indeed, if we believe that we
can avoid the principles of quantum mechanics and replace them by what
is called "classical change of signature", what should it be true only
for the background solution and not for the perturbed one?) it will lead to
conflicts with observations. The absence of gravitational waves in this model
is due to the restrictiveness of the junction condition $K_{ij}=0$.
\par
It is worth noticing that in the context of the classical change of signature,
an alternative proposal for the junction conditions has been advocated
\cite{ref4,ref6,ref15,ref16,ref22}. For this proposal, the matching conditions
require the continuity only of the second fundamental form. In our model, this
means the continuity of $\dot{h}_{ij}$, or the continuity of $\frac{{\rm d}\nu
_N}{{\rm d}t}$. Therefore, the regular solution of $\nu _N$ will be able to
cross the surface of change $\Sigma $ leading to a non-vanishing amplitude for
the gravitational waves in the inflationary era.
\par
However, despite this possibility, it seems that a theory including both
general relativity and quantum mechanics is more likely to provide a
satisfactory description of the primordial Universe.

\section{Acknowledgments}
This work was initiated in discussions with R. Kerner; his invaluable help and
constant encouragement are here gratefully acknowledged. I would like to
express my gratitude to L. P. Grishchuk for several enlightening discussions
concerning this paper. I would like also to thank G. Gibbons, B. Linet and L.
Bel for usefull discussions.
\par
The work was supported by a grant of the Minist\`ere de la Recherche et de
l'Enseignement Superieur (France).

\newpage

{\bf Figures captions:}

\vspace{1cm}

Figure 1: Different system of coordinates for the Riemannian region.

\vspace{1cm}

Figure 2: Contours for the propagator.

\end{document}